\documentclass{article}

\usepackage{arxiv}

\usepackage[utf8]{inputenc}     
\usepackage[T1]{fontenc}        
\usepackage{hyperref}           
\usepackage{url}                
\usepackage{booktabs}           
\usepackage{amsmath, amsfonts, amssymb}
\usepackage{nicefrac}           
\usepackage{microtype}          
\usepackage{graphicx}
\usepackage{subcaption}
\usepackage{natbib}
\usepackage{doi}
\usepackage{color}
\usepackage{xcolor}

\title{Beyond the Mean: A Two-Stage GAMLSS Framework for Physical Activity Bout Patterns and Survival in COPD}

\renewcommand{\shorttitle}{Two-Stage GAMLSS for Physical Activity Bouts and Survival in COPD}

\hypersetup{
pdftitle={Beyond the Mean: A Two-Stage GAMLSS Framework for Physical Activity Bout Patterns and Survival in COPD},
pdfsubject={stat.ME, stat.AP},
pdfauthor={Irantzu Barrio, Crist\'obal Esteban, Thomas Kneib},
pdfkeywords={GAMLSS, distributional regression, survival analysis, COPD, physical activity patterns},
}

\author{%
  \href{https://orcid.org/0000-0000-0000-0000}{\includegraphics[scale=0.06]{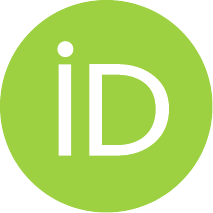}\hspace{1mm}Irantzu Barrio} \\
  Department of Mathematics \\
  University of the Basque Country UPV/EHU, Spain \\
  BCAM -- Basque Center for Applied Mathematics, Spain \\
  \texttt{irantzu.barrio@ehu.eus} \\
  \And
  Crist\'obal Esteban \\
  Respiratory Unit \\
  Galdakao University Hospital, Spain \\
      \texttt{cristobal.est@gmail.com} \\
  \And
  \href{https://orcid.org/0000-0003-3390-0972}{\includegraphics[scale=0.06]{orcid.pdf}\hspace{1mm}Thomas Kneib} \\
  Chair of Statistics and Campus Institute Data Science \\
  University of G\"ottingen, Germany \\
    \texttt{tkneib@uni-goettingen.de} \\
}

\date{}

\begin{document}
\maketitle

\begin{abstract}
Physical activity patterns in chronic obstructive pulmonary disease (COPD) are typically summarised using simple metrics such as total daily activity or mean time spent in moderate-to-vigorous physical activity. In many studies, this activity is further characterized through physical activity bouts, defined as periods of at least 10 consecutive minutes above 3 metabolic equivalent tasks (METs). However, such summaries overlook the highly skewed, heterogeneous and patient-specific nature of bout-related variables, which may contain clinically relevant information. We propose a distributional modelling approach for accelerometer-derived bout-related physical activity outcomes based on Generalised Additive Models for Location, Scale and Shape (GAMLSS). The framework models multiple parameters of the outcome distributions, including location, scale, and the probability of observing no bouts, allowing the analysis to extend beyond average levels of activity. In a second stage, the estimated distributional parameters are incorporated into a survival analysis to investigate whether information beyond the mean is associated with patient prognosis. The proposed framework is illustrated using accelerometer data from a cohort of 512 COPD patients and considering bout-related outcomes such as minutes accumulated in bouts per day, total number of bouts, and area under MET in bouts per day. The resulting distributional parameters were significantly associated with mortality and provided information beyond average activity levels, particularly through differences in variability across patients. These findings suggest that information contained in distributional parameters beyond the mean may be relevant for understanding the relationship between physical activity patterns and survival.
\end{abstract}


\section{Introduction}

In the general population, physical activity (PA) is considered a key aspect of a healthy lifestyle, having a direct relationship with mortality \citep{ekelund2026deaths}.  In patients with Chronic Obstructive Pulmonary Disease (COPD), a paradigmatic chronic respiratory disease, PA has been observed to be one of the key factors associated with health related quality of life (HRQoL) and furthermore with the change over time of HRQoL \citep{esteban2010impact, waschki2015disease, esteban2020predictive} and survival. Indeed, PA has been identified as one of the most important factors related to mortality in this population \citep{waschki2011physical}.

Most studies in COPD have evaluated PA through measures of the total volume of activity accumulated over a given period.  However, PA behaviour is highly heterogeneous across patients, reflecting differences in disease severity, respiratory function, comorbidity burden, and other clinical and demographic characteristics \citep{gimeno2014determinants, llamas2023physical, vasquez2025role}. As a consequence, patients with similar total amounts of activity may differ substantially in the way this activity is accumulated throughout daily life. In this context, the evaluation of PA bouts \citep{donaire2013} and their characteristics provides an opportunity to describe activity patterns in greater detail and may help identify patient profiles associated with different clinical outcomes.

However, even when focusing on PA bout-related outcomes, most analyses continue to rely on summary measures that describe only the average behaviour. Although such measures are clinically meaningful and easy to interpret, they may overlook relevant differences in variability, asymmetry, or the occurrence of inactivity. Furthermore, bout-related outcomes derived from accelerometer data frequently exhibit skewed distributions, excess zeros, and substantial between-patient heterogeneity, making conventional mean-based approaches potentially inadequate for capturing their full complexity. Statistical methods capable of modelling multiple aspects of the outcome distribution may therefore provide a more complete description of bout-related variables.

Generalized Additive Models for Location, Scale and Shape (GAMLSS) offer a flexible framework for modelling several parameters of the outcome distribution simultaneously as functions of explanatory variables \citep{stasinopoulos2024generalized}. By extending the analysis beyond the mean, these models make it possible to investigate how clinical and demographic characteristics are associated not only with the expected level of an outcome but also with its variability and other distributional parameters.

Although GAMLSS have been successfully applied in a variety of clinical settings, including the analysis of treatment effects in cardiovascular clinical trials \citep{heller2022distributional} and the development of normative growth charts and reference trajectories in neuroimaging studies \citep{bethlehem2022brain}, their use to study bout-related outcomes remains limited. In particular, little attention has been paid to modelling different distributional parameters of these outcomes and to understanding how clinical and demographic characteristics influence not only their expected values but also their variability and other aspects of their distributions.

In addition, beyond the characterization of bout-related PA outcomes, the relationship between distributional parameters and clinically important outcomes has received little attention. In particular, it remains unclear whether parameters beyond the mean may provide additional information for understanding patient prognosis.

Therefore, the aim of this work was to propose a distributional modelling framework for bout-related physical activity outcomes derived from accelerometer data in patients with COPD. The proposed approach combines GAMLSS models to characterise different distributional parameters of the outcomes and a subsequent survival analysis based on the resulting estimated distributional parameters. Through this framework, we illustrate how modelling parameters beyond the mean can provide additional insights into physical activity behaviour and their relationship with survival.

The remainder of the paper is organised as follows. Section 2 introduces the COPD data set that motivated this work. Section 3 presents the proposed modelling approach and the associated statistical methodology. Section 4 illustrates the proposed approach through its application to the COPD data set. Finally, Section 5 concludes with a discussion of the main findings, limitations and directions for further research.

\section{COPD cohort and physical activity data}

We analyse a prospective cohort of 512 COPD patients from the Galdakao University Hospital (Spain). Patients were recruited consecutively if they had a confirmed diagnosis of COPD for at least six months and had remained clinically stable for a minimum of six weeks. Individuals were excluded if they had asthma, other significant respiratory disorders, or psychiatric or neurological conditions that could interfere with study participation. The study protocol received approval from the hospital's ethics and research committees (16/2014). All potential participants were fully informed about the study, and written informed consent was obtained from all enrolled patients.

Clinical and sociodemographic variables were collected at baseline, and up to three annual follow-up visits were conducted. Comorbidities were
summarized using the Charlson comorbidity index (CCI) \citep{charlson1987new}. Pulmonary function tests included forced expiratory
volume in one second in percentage (FEV$_1\%$), and measurements of carbon monoxide diffusing capacity (DLCO$\%$), residual volume (RV$\%$), total lung capacity (TLC$\%$), and maximal inspiratory pressure (PIM$\%$), all expressed as percentages of theoretical predicted values. These tests were performed in accordance with the standards of the European Respiratory Society. For reference values, we considered those of the European Respiratory Society \citep{stanojevic2017official}.
The physical capacity of patients was evaluated by the 6-minute walk test's (6MWT) total distance in meters. Physical activity (PA) was measured using an accelerometer (SenseWear Pro 3 Body Media Inc, Pittsburgh, USA). Participants kept the armband on over 7 consecutive days after recruitment, at all times except when performing daily personal hygiene. All the variables used in this work were those measured at baseline.

\begin{figure}[!ht]\centering
\includegraphics[width=12cm]{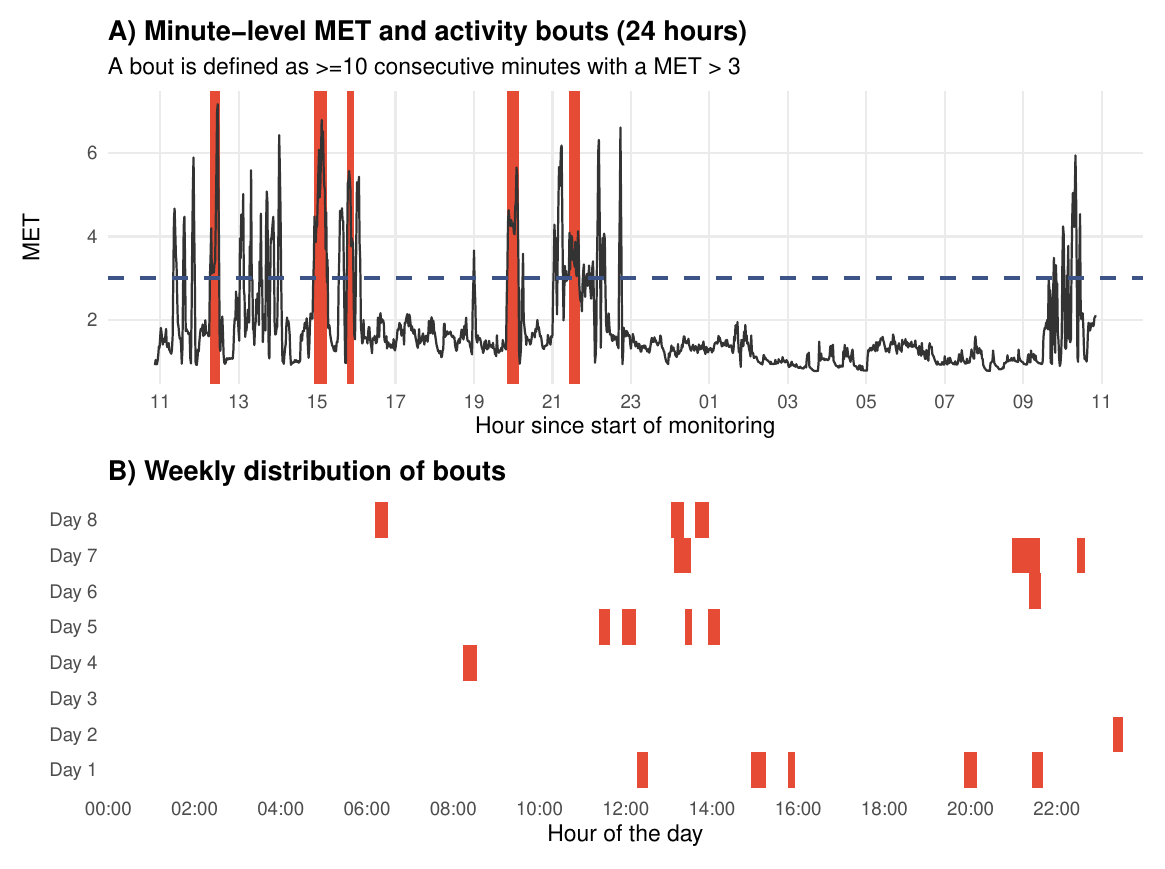}
\caption{\label{barrio:fig_bout_representation} Illustrative representation of the temporal evolution of MET and weekly distribution of bouts for one illustrative patient. Panel A) represents the temporal evolution of MET during the first 24 hours with the accelerometer with the orange-shaded area representing a bout. Panel B) shows the distribution of bouts over the 7 days using the accelerometer.}
\end{figure}

Of the 512 patients, 472 provided valid accelerometer data. Following previous COPD research, PA bouts were defined as periods of $\ge 10$ minutes above 3 metabolic equivalent tasks (METs), corresponding to moderate-to-vigorous physical activity (MVPA) \citep{ainsworth20112011, donaire2013}. Of 472 patients, 432 (91.5\%) had at least one bout during the time they wore the accelerometer. Figure~\ref{barrio:fig_bout_representation} shows a representation of the temporal evolution of MET during the first 24 hours, together with the total number of bouts distributed along the 7 days for one illustrative patient. We accounted for the total time patients expend in bouts (minutes), time patients expend in bouts per day (minutes), total number of bouts, number of bouts per day, the total area under MET in bouts and the total area under MET in bouts per day. For ease of presentation, hereafter we refer to these variables using the following shorthand notation: total time spent in bouts ($\text{Time}_b$), time spent in bouts per day ($\text{Time}_{bd}$), total number of bouts ($\text{Count}_b$), number of bouts per day ($\text{Count}_{bd}$), total area under MET in bouts ($\text{METarea}_b$), and total area under MET in bouts per day ($\text{METarea}_{bd}$).  Table~\ref{tab:descriptive} provides a description of the variables considered for this study.

\begin{table}[htbp]
\centering
\caption{Descriptive characteristics of the sample with accelerometer data $n=472$}
\label{tab:descriptive}
\begin{tabular}{lrrrrr}
\hline
\textbf{Variable} & \textbf{Median} & \textbf{Q1--Q3} & \textbf{Mean} & \textbf{SD} & \textbf{Missing} \\
\hline
\multicolumn{6}{l}{\textit{Continuous variables}} \\
Age (years) & 65 & 58--71 & 64.41 & 8.78 &  \\
BMI, kg/m$^2$ & 27 & 23.97--30.18 & 27.45 & 5.11 & \\
DLCO$\%$ & 60.9 & 46.02--76.5 & 61.72 & 20.89 & 2 \\
PIM$\%$ & 58.4 & 44.33--75.72 & 60.18 & 22.43 &  \\
RV$\%$ & 165.5 & 138.65-195 & 170.43 & 50.62 & 1 \\
TLC$\%$ & 117.7 & 106.55-130.7 & 119.02 & 21.18 & 1 \\
6MWT (meters) & 485.5 & 416.75--549 & 473.42 & 108.57 &  \\
Time with accelerometer (days) & 7 & 6.94--7.01 & 6.96 & 0.24 &  \\
$\text{Time}_{b}$ & 232 & 69-542.25 & 422.25 & 524.92 & 0\\
$\text{Time}_{bd}$ & 33.17 & 9.85--77.41 & 60.75 & 75.48 & \\
$\text{Count}_{b}$ & 13.5 & 5--28 & 19.48 & 20.23 &  \\
$\text{Count}_{bd}$ & 1.93 & 0.71-3.99 & 2.8 & 2.9 & 0\\
$\text{METarea}_{b}$ & 1026.97 & 320.39-2577.72 & 2009.43 & 2747.58 & 0\\
$\text{METarea}_{bd}$ & 146 & 45.77-367.98 & 289.34 & 395.68 & 0 \\
Follow up time (days) & 1733.5 & 1088--1873.5 & 1488.51 & 519.41 &  \\
\hline
 & \textbf{$n$} & \textbf{$\%$} & \multicolumn{2}{l}{} & \\
\hline
\multicolumn{6}{l}{\textit{Categorical variables}} \\
Sex: Men & 347  & 73.52 &  & &  \\
Sex: Female & 125  & 26.48 &  & &  \\
Previous hospitalizations: 0 & 314  & 66.53 &  & &  \\
Previous hospitalizations: >=1 & 158  & 33.47 &  & &  \\
Charlson index: 1 & 260 & 55.08 &  &  & \\
Charlson index: 2 & 122  & 25.85 &  &  & \\
Charlson index: >2 & 90  & 19.07 &  &  & \\
FEV$_1\%$: >=50 & 296  & 62.71 &  &  & \\
FEV$_1\%$: <50 & 176  &37.29  &  &  & \\
BMI: <30 & 348  &73.73  &  &  & \\
BMI: >=30 & 124 & 26.27 &  &  & \\
Mortality along the study: Yes & 78 & 16.53 &  &  & \\
\hline
\end{tabular}

\vspace{0.3cm}
\footnotesize{
Continuous variables are summarized using median, interquartile range (Q1--Q3), mean, and standard deviation (SD).
Categorical variables are presented as number (percentage). BMI: body mass index.
}
\end{table}

\begin{figure}[htbp]
\centering

\begin{subfigure}[b]{0.32\textwidth}
    \centering
    \includegraphics[width=\textwidth]{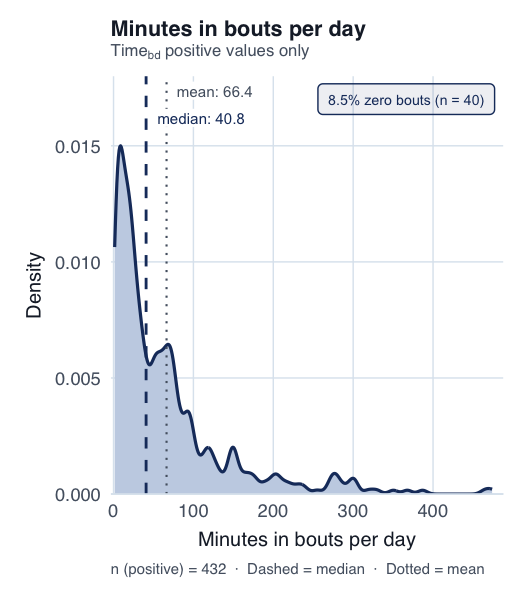}
    \caption{}
\end{subfigure}
\hfill
\begin{subfigure}[b]{0.32\textwidth}
    \centering
    \includegraphics[width=\textwidth]{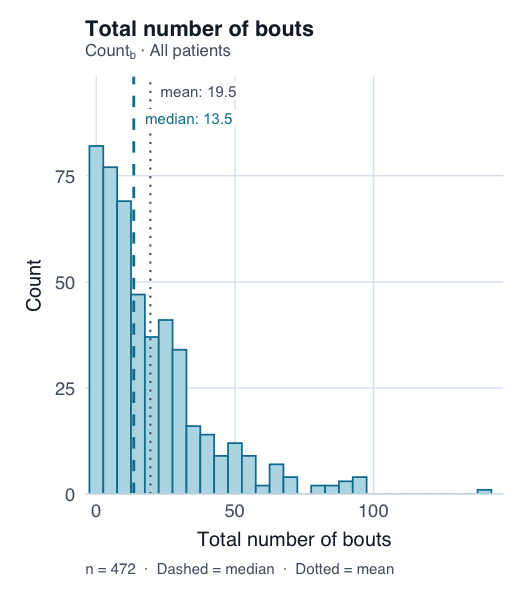}
    \caption{}
\end{subfigure}
\hfill
\begin{subfigure}[b]{0.32\textwidth}
    \centering
    \includegraphics[width=\textwidth]{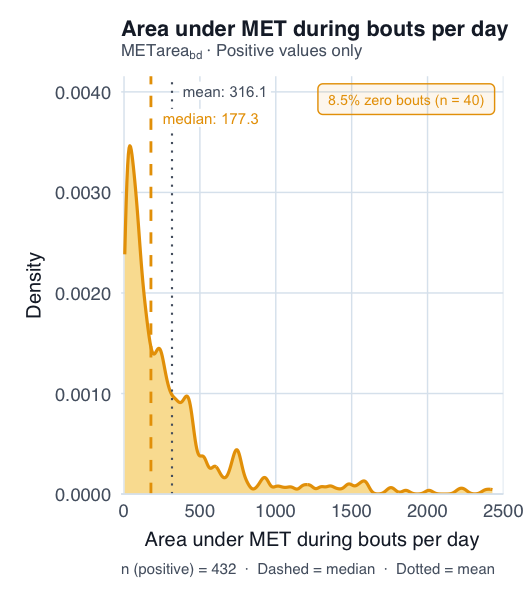}
    \caption{}
\end{subfigure}

\caption{Distribution of the Physical Activity outcomes. From left to right, (a) total minutes in bouts per day; (b) total number of bouts; (c) total area under MET during bouts per day.}
\label{barrio:fig_dist}
\end{figure}

We focus on the variables $\text{Time}_{bd}$, $\text{Count}_{b}$ and $\text{METarea}_{bd}$ which exhibit substantial right skewness and variability across patients, and provide an overall representation of the patient's PA (Figure \ref{barrio:fig_dist}). The distributional characteristics of these variables motivate the use of GAMLSS for flexible distributional modelling.

\section{Methods}
In this section, we present the proposed modelling framework for accelerometer-derived bout-related physical activity outcomes. In Section \ref{subsec:methods_overview} we present an overview of the proposed modelling framework, while in Section \ref{subsec:methods_gamlss} and Section \ref{subsec:methods_survival} we provide the specific details of GAMLSS and Survival modelling, respectively.

\subsection{Overview of the Proposed Approach}\label{subsec:methods_overview}
\begin{figure}[!ht]\centering
\includegraphics[width=12cm]{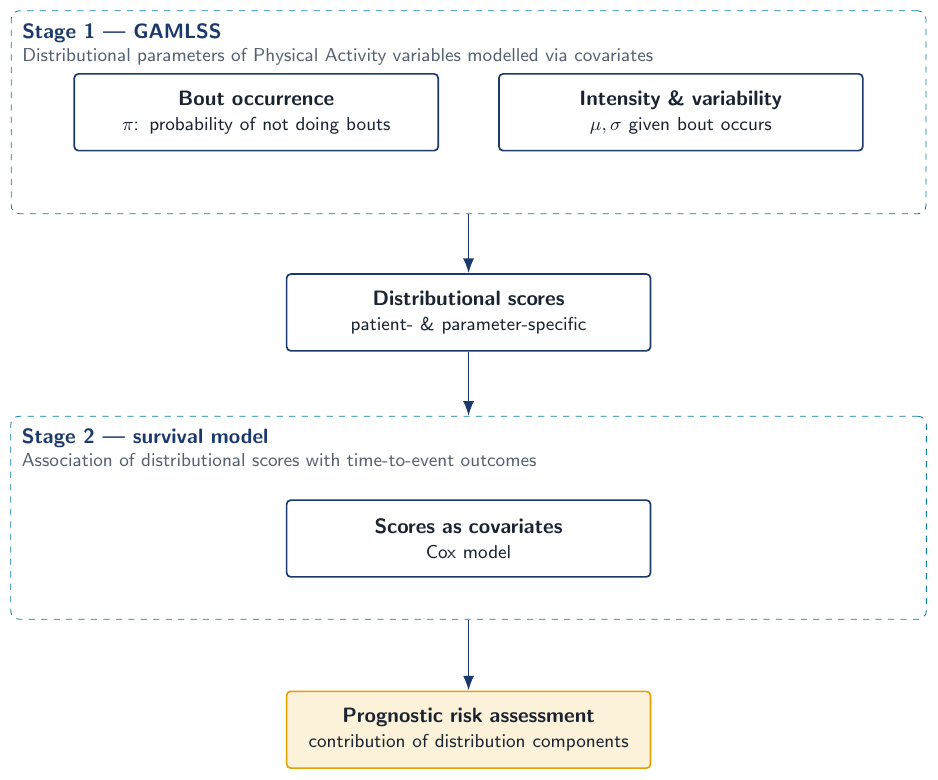}
\caption{\label{fig:framework} Graphical overview of the proposed approach.}
\end{figure}

The proposed modelling approach consists of two stages. In the first stage, variables of interest are modelled using a Generalised Additive Model for Location, Scale and Shape (GAMLSS), allowing multiple parameters of the response distribution - such as location and scale - to be modelled as functions of covariates. To account for the excess of zero values (zero bouts), a zero-adjusted formulation is considered, distinguishing between the probability of engaging in bouts and, conditional on doing bouts, the intensity and variability of the response.  This yields patient‑ and parameter‑specific scores that characterise key aspects of the entire distribution, rather than relying on single summary measures. These distributional scores capture not only the expected value of the variable but also its variability and heterogeneity across individuals. In the second stage, the estimated distributional scores are incorporated as covariates in a survival model to assess their association with time‑to‑event outcomes. This two‑stage framework separates the modelling of complex, highly skewed response distributions from the survival analysis, while retaining clinically interpretable summaries of different distributional components. It is important to note that these two stages operate on different time scales: the distributional parameters are estimated from accelerometer measurements collected during a 7-days monitoring period at baseline, whereas survival is evaluated over a 4-years follow-up period. In this way, the approach enables the evaluation of how distinct aspects of the underlying distribution contribute to prognostic risk. A graphical overview of the proposed approach is shown in Figure~\ref{fig:framework}.

\subsection{GAMLSS}\label{subsec:methods_gamlss}
Let $Y$ be a response variable with a probability distribution $\mathcal{D}$ defined by $K$ parameters $\theta^k$ $k \in \{1,\ldots,K \}$. Thus, for each individual $i \in \{1,\ldots,n \}$, \(Y_{i}\sim \mathcal{D}(\theta _{i}^1,\theta _{i}^2,\dots ,\theta _{i}^K)\). Let $X_{i1}, \ldots, X_{iM}$ be a set of $M$ predictor variables for each individual $i$. For ease of notation, we assume these are either continuous or dummy predictor variables. Then, each parameter $\theta_{i}^k$ with $k \in \{1,\ldots,K \}$ and $i \in \{1,\ldots,n \}$ is linked to a set of predictor variables via $\theta_{i}^k = g_k(\eta_{i}^k)$, where
$$
\eta_{i}^k \;=\; \beta_0^{k} \;+\; \sum_{j=1}^{J}\beta_j^{k}\,X_{ij} \;+\; \sum_{m=J+1}^{M} f_m^{k}(X_{im}),
$$
with $\beta_0^{k}$ an intercept, $\beta_j^{k}$ parametric effects, and $f_m^{k}(\cdot)$ smooth functions for selected continuous covariates. Parameters are estimated by maximum likelihood using the iterative RS (Rigby -- Stasinopoulos) backfitting algorithm \citep{rigby2005generalized}, yielding the individual fitted scores $\widehat{\eta}_{i}^k$. These fitted predictors provide low-dimensional, interpretable summaries of different characteristics of the response distribution at the individual level.

In this work, we focus primarily on modelling the location ($\theta^1=\mu$) and scale ($\theta^2=\sigma$) parameters of the response distribution. The location parameter captures the expected value of the outcome, while the scale parameter reflects intra-individual variability and heterogeneity. Both components may carry complementary prognostic information, particularly in settings where response distributions are highly skewed or overdispersed. In addition, we consider continuous or semi-continuous distributions supported on the positive real line and unbounded discrete distributions, which naturally arise in the analysis of physical activity variables such as accumulated time, area under MET or bout counts. Specifically, we use the Gamma and inverse Gaussian distributions for modelling $\text{Time}_{bd}$ and $\text{METarea}_{bd}$, and the Poisson and negative binomial for modelling  $\text{Count}_{b}$ \citep{stasinopoulos2024generalized}.

For semi-continuous outcomes with an excess of zero values, such as $\text{Time}_{bd}$ or $\text{METarea}_{bd}$, which by definition are zero for those patients without any bout, a zero-adjusted formulation is adopted. This allows separating the probability of engaging in bouts from the distribution of positive values, for which location and scale parameters are modelled.
Therefore, the response distribution ($\text{Time}_{bd}$ or $\text{METarea}_{bd}$) can be modelled as a mixture of a point mass at zero and a continuous distribution for positive values. Specifically, let $\nu_i = \Pr(Y_i = 0)$ denote the probability of not doing bouts for individual $i$. Then, the response variable is modelled as
\[
Y_i \sim
\begin{cases}
0, & \text{with probability } \nu_i, \\
\mathcal{D}(\mu_i, \sigma_i), & \text{with probability } 1- \nu_i,
\end{cases}
\]
where the distribution $\mathcal{D}$ stands for either the Gamma or inverse gaussian distribution and is parametrised by mean $\mu_i$ and scale $\sigma_i$, and represents the distribution of positive responses, i.e., $Y_i \mid Y_i > 0$.

To facilitate interpretation and comparability in the subsequent survival analysis, we express variability in terms of the standard deviation of the fitted response distribution rather than the raw scale parameter $\sigma$. While $\sigma$ is distribution-specific and may not have a direct interpretation on the outcome scale, the standard deviation provides a common and clinically interpretable measure of dispersion. For each fitted model, the standard deviation is computed as a deterministic function of the estimated distributional parameters.
Under the zero-adjusted formulation, we define the variability measure conditionally on positive responses. For instance, when $\mathcal{D}=\text{Gamma}$ is considered, parametrised by mean $\mu$ and scale $\sigma$ in the \texttt{GA()} function used by the package \texttt{gamlss} in \texttt{R}, the variance of the positive component is given by $\mathrm{Var}(Y \mid Y>0)=\mu^2\sigma^2$, and therefore the corresponding standard deviation is $\mathrm{Sd}(Y \mid Y>0)=\mu\sigma$. For count outcomes modelled using a negative binomial distribution (function \texttt{NBI()} in \texttt{gamlss.dist}), the variance is $\mathrm{Var}(Y)=\mu(1+\sigma\mu)$, yielding a standard deviation of $\mathrm{Sd}(Y)=\sqrt{\mu(1+\sigma\mu)}$ \citep{rigby2019distributions}.

Taken together, these quantities provide a multi-component representation of individual behaviour. Therefore, for semi-continuous outcomes, this representation comprises three components: the probability of engaging in bouts, the expected intensity of the response (conditional on positive values), and the associated intra-individual variability (also conditional on positive values). In contrast, for count outcomes, the characterisation is based on two components, capturing the expected number of bouts and its variability.

This representation moves beyond traditional single-summary measures by incorporating both participation and distributional parameters of the response. In the second stage of the analysis, the corresponding distributional parameters are used as covariates in the survival model, enabling the assessment of their joint and potentially complementary effects on the time-to-event.

\subsection{Survival Modeling}\label{subsec:methods_survival}
Let $T$ denote the time to event and $C$ the censoring time, with event indicator $\delta=\mathbb{I}(T \le C)$ and observed time $T^*=\min(T,C)$. To evaluate the prognostic relevance of the distributional characteristics estimated in Section~\ref{subsec:methods_gamlss}, we incorporate the GAMLSS-derived summaries of the response distribution as covariates in a Cox proportional hazards model.

For semi-continuous outcomes with excess zeros, such as $\text{Time}_{bd}$ or $\text{METarea}_{bd}$, we consider three components:  the probability of not doing bouts, denoted by $\widehat{\nu}$, and, conditional on doing bouts, the estimated mean and standard deviation of the positive response distribution, denoted by $\widehat{\mu}$ and $\widehat{\mathrm{Sd}}$, respectively. For count outcomes such as $\text{Count}_{b}$, the full distribution is modelled directly, and therefore only the estimated mean and standard deviation are considered.

These quantities summarise distinct and complementary aspects of the response, namely doing or not doing bouts, expected intensity, and intra-individual variability. Then, the hazard function is specified as
\[
h(t \mid \widehat{\nu}, \widehat{\mu}, \widehat{\mathrm{Sd}})
=
h_0(t)\exp\left(
\omega_\nu \widehat{\nu}
+
\omega_\mu \widehat{\mu}
+
\omega_{\mathrm{Sd}} \widehat{\mathrm{Sd}}
\right),
\]
for semi-continuous outcomes, and analogously without the $\widehat{\nu}$ term for count outcomes, where $h_0(t)$ denotes the baseline hazard function and $\omega_\nu$ , $\omega_\mu$ and $\omega_{\mathrm{Sd}}$ quantify the association between survival risk and the estimated probability of not doing bouts, estimated mean and standard deviation of the response distribution, respectively.

The probability component is included on its natural scale to facilitate interpretation; alternative parameterisations based on the linear predictor yielded equivalent inferential conclusions.

We first fit separate Cox models for each PA measure, incorporating the corresponding distributional components. Later, in a final multivariable model, components derived from $\text{Time}_{bd}$, $\text{METarea}_{bd}$ and $\text{Count}_{b}$ are jointly included to assess their combined prognostic value. Finally the models discrimination ability is measured by means of Harrell's concordance index (c-index, \cite{harrell1996multivariable}) provided by the function \texttt{concordance()} in the \texttt{survival} package \citep{therneau2025survival}.

\section{Results}

We modelled the location and scale parameters physical activity variables that best reflect activity patterns in patients with COPD, namely $\text{Time}_{bd}$, $\text{METarea}_{bd}$ and $\text{Count}_{b}$. For semi-continuous variables, we also modelled the probability of engaging in bouts. The distributional modelling results for these variables are presented in Sections~\ref{subsec:results_minutes}, ~\ref{subsec:results_areaMET}, and ~\ref{subsec:results_bouts}, respectively. Section~\ref{subsec:results_survival} then links these distributional characteristics to prognosis by incorporating the estimated distributional components of the outcomes as covariates in the survival analysis. Finally, in Section~\ref{subsec:results_interpretation} we interpret the joint effect of the distributional components by the illustration of different patient profiles.

\subsection{Modelling \textit{minutes in bouts per day}} \label{subsec:results_minutes}
When comparing the fits among the Gamma and inverse Gaussian distribution, the zero-adjusted Gamma (\texttt{ZAGA()}) provided the best fit to model the variable $\text{Time}_{bd}$.  We modelled $\mu_i$ and $\sigma_i$, $i\in \{1,\ldots,n\} $, with \textit{log} links and $\nu_i$ with \textit{logit} link. Candidate predictors included age, sex, body mass index (BMI), CCI, number of previous hospitalizations, diffusing capacity of the lung for carbon monoxide (DLCO\%), forced expiratory volume in one second (FEV1\%), maximal inspiratory pressure (PIM\%), residual volume (RV$\%$), total lung capacity (TLC$\%$), and 6-minute walking test (6MWT). The best predictor variables were selected using the minimum AIC criteria. The obtained results are shown in Table~\ref{barrio:tab_minutes_ZAGA}. All the variables introduced as continuous had a linear effect.

The ZAGA model showed that the amount of time accumulated in bouts per day was associated with several demographic and clinical characteristics through different distributional parameters. In the location submodel ($\mu$), older age was associated with lower expected minutes in bouts, while higher DLCO$\%$ and PIM$\%$ were associated with higher values. Female sex showed a borderline negative association ($p=0.061$). Greater comorbidity burden (CCI = 2 and especially CCI > 2) and having at least one previous hospitalization were also associated with lower expected time in bouts. In the scale submodel ($\sigma$), patients with FEV1$\% < 50$ and those achieving $\geq 460$ meters in the 6-minute walking test exhibited lower variability in bout time. Regarding the zero-adjustment parameter ($\nu$), which models the probability of doing no bouts, older age and obesity (BMI $\geq 30$) were associated with a higher probability of zero values, whereas higher  DLCO$\%$  was associated with a lower probability of not doing bouts.

\begin{table}[!ht]
\centering
\caption{GAMLSS results for location ($\mu$), scale ($\sigma$), and zero-adjustment ($\nu$) of the \textit{minutes accumulated in bouts per day} with log (for $\mu,\sigma$) and logit (for $\nu$) link functions. The $\nu$ parameter captures the probability of a zero response (no bouts), thereby complementing the information provided by the location and scale parameters.}
\label{barrio:tab_minutes_ZAGA}
\begin{tabular}{lrrrrrrrrr}
\toprule
&
\multicolumn{3}{c}{Location ($\mu$)} &
\multicolumn{3}{c}{Scale ($\sigma$)} &
\multicolumn{3}{c}{Zero-adjustment ($\nu$)} \\
\cmidrule(lr){2-4} \cmidrule(lr){5-7} \cmidrule(lr){8-10}
& Est. & Std. Err & p-value
& Est. & Std. Err & p-value
& Est. & Std. Err & p-value \\
\midrule
(Intercept)
& 4.308 & 0.428 & $<0.001$
& 0.198 & 0.061 & 0.001
& -3.311 & 1.533 & 0.031 \\

Age
& -0.016 & 0.006 & 0.004
& --- & --- & ---
& 0.056 & 0.023 & 0.015 \\

Sex - Female
& -0.222 & 0.118 & 0.061
& --- & --- & ---
& --- & --- & --- \\

CCI $= 2$
& -0.398 & 0.113 & $<0.001$
& --- & --- & ---
& --- & --- & --- \\

CCI $> 2$
& -0.585 & 0.137 & $<0.001$
& --- & --- & ---
& --- & --- & --- \\

Prev. hosp $\ge 1$
& -0.239 & 0.104 & 0.022
& --- & --- & ---
& --- & --- & --- \\

DLCO$\%$
& 0.013 & 0.002 & $<0.001$
& --- & --- & ---
& -0.062 & 0.011 & $<0.001$ \\

PIM$\%$
& 0.005 & 0.002 & 0.020
& --- & --- & ---
& --- & --- & --- \\

FEV1$\%$ $<50$
& --- & --- & ---
& -0.147 & 0.067 & 0.028
& --- & --- & --- \\

6MWT $\ge 460$
& --- & --- & ---
& -0.316 & 0.066 & $<0.001$
& --- & --- & --- \\

BMI $\ge 30$
& --- & --- & ---
& --- & --- & ---
& 1.690 & 0.395 & $<0.001$ \\

\bottomrule
\end{tabular}
\end{table}

\subsection{Modelling \textit{area under MET per day}} \label{subsec:results_areaMET}

When comparing the fits among the Gamma and inverse Gaussian distribution, the zero-adjusted Gamma (\texttt{ZAGA()}) provided the best fit to model the variable $\text{METarea}_{bd}$.  We modelled $\mu_i$ and $\sigma_i$, $i\in \{1,\ldots,n\} $, with \textit{log} links and $\nu_i$ with \textit{logit} link. Candidate predictors included age, sex, body mass index (BMI), CCI, number of previous hospitalizations, diffusing capacity of the lung for carbon monoxide (DLCO\%), forced expiratory volume in one second (FEV1\%), maximal inspiratory pressure (PIM\%), residual volume (RV$\%$), total lung capacity (TLC$\%$), and 6-minute walking test (6MWT). The best predictor variables were selected using the minimum AIC criteria. The obtained results are shown in Table~\ref{tab:METarea_ZAGA}. All the variables introduced as continuous had a linear effect.

Results for the area under MET per day were largely consistent with those observed for minutes accumulated in bouts. The expected value ($\mu$) decreased with age, increasing comorbidity burden, and obesity, while higher DLCO\% and PIM\% were associated with larger values. Sex was not significantly associated with the outcome. For the scale parameter ($\sigma$), lower variability was observed in patients with FEV1$\% < 50$, higher 6MWT performance ($\geq 460$), and higher TLC\%, whereas RV\% was positively associated with variability. The zero-adjustment component ($\nu$) was identical to that reported for minutes accumulated in bouts.

\begin{table}[!ht]
\centering
\caption{GAMLSS results for location ($\mu$), scale ($\sigma$), and zero-adjustment ($\nu$) of the \textit{area under MET per day} with log (for $\mu,\sigma$) and logit (for $\nu$) link functions. The $\nu$ parameter captures the probability of a zero response (no bouts), thereby complementing the information provided by the location and scale parameters.}
\label{tab:METarea_ZAGA}
\begin{tabular}{lrrrrrrrrr}
\toprule
&
\multicolumn{3}{c}{Location ($\mu$)} &
\multicolumn{3}{c}{Scale ($\sigma$)} &
\multicolumn{3}{c}{Zero-adjustment ($\nu$)} \\
\cmidrule(lr){2-4} \cmidrule(lr){5-7} \cmidrule(lr){8-10}
& Est. & Std. Err & p-value
& Est. & Std. Err & p-value
& Est. & Std. Err & p-value \\
\midrule
(Intercept)
& 6.277 & 0.443 & $<0.001$
& 0.484 & 0.201 & 0.016
& -3.311 & 1.533 & 0.031 \\

Age
& -0.024 & 0.006 & $<0.001$
& --- & --- & ---
& 0.056 & 0.023 & 0.015 \\

Sex - Female
& -0.186 & 0.117 & 0.111
& --- & --- & ---
& --- & --- & --- \\

CCI $= 2$
& -0.401 & 0.112 & $<0.001$
& --- & --- & ---
& --- & --- & --- \\

CCI $> 2$
& -0.608 & 0.132 & $<0.001$
& --- & --- & ---
& --- & --- & --- \\

BMI $\ge 30$
& -0.873 & 0.109 & $<0.001$
& --- & --- & ---
& 1.690 & 0.395 & $<0.001$ \\

DLCO$\%$
& 0.016 & 0.002 & $<0.001$
& --- & --- & ---
& -0.062 & 0.011 & $<0.001$ \\

PIM$\%$
& 0.005 & 0.002 & 0.034
& --- & --- & ---
& --- & --- & --- \\

FEV1$\%$ $<50$
& --- & --- & ---
& -0.242 & 0.076 & 0.002
& --- & --- & --- \\

6MWT $\ge 460$
& --- & --- & ---
& -0.312 & 0.065 & $<0.001$
& --- & --- & --- \\

RV$\%$
& --- & --- & ---
& 0.004 & 0.001 & 0.003
& --- & --- & --- \\

TLC$\%$
& --- & --- & ---
& -0.008 & 0.003 & 0.010
& --- & --- & --- \\

\bottomrule
\end{tabular}
\end{table}

\subsection{Modelling \textit{total number of bouts}}\label{subsec:results_bouts}
When comparing the fits obtained with the Poisson and the negative binomial distribution, the latter provided the best fit to model the variable $\text{Count}_{b}$. We did not observe the need to use a zero-inflated distribution. Instead we considered the total time wearing the accelerometer as the exposure variable, to model the rate of number of bouts per day, although since the exposure time was almost constant (IQR 6.94--7.01) for all patients it did not influence the results obtained. We modelled $\mu_i$ and $\sigma_i$ with log links. Candidate predictors included age, sex, BMI, CCI, number of previous hospitalizations, DLCO\%, FEV1\%, PIM\%, RV$\%$, TLC$\%$, and 6MWT. The best predictor variables were selected using the minimum AIC criteria. The obtained results are shown in Table~\ref{barrio:tab_count_NBI}. All the variables introduced as continuous had a linear effect.

In the negative binomial model for the total number of bouts, the expected count ($\mu$) was negatively associated with age, comorbidity burden, obesity, and previous hospitalizations, and positively associated with DLCO\%, PIM\%, and achieving $\geq 460$ meters in the 6MWT. The scale parameter ($\sigma$) was higher among obese individuals, whereas higher DLCO\% and better 6MWT performance were associated with lower dispersion; a similar trend was observed for FEV1$\% < 50$, although the association was only borderline significant ($p=0.056$).

\begin{table}[!ht]
\centering
\caption{GAMLSS results for location ($\mu$) and scale ($\sigma$) of the \textit{total number of bouts} with log link functions.}
\label{barrio:tab_count_NBI}
\begin{tabular}{lrrrrrr}
\toprule
&
\multicolumn{3}{c}{Location ($\mu$)} &
\multicolumn{3}{c}{Scale ($\sigma$)} \\
\cmidrule(lr){2-4} \cmidrule(lr){5-7}
& Est. & Std. Err & p-value
& Est. & Std. Err & p-value \\
\midrule
(Intercept)
& 0.638 & 0.348 & 0.068
& 1.327 & 0.273 & $<0.001$ \\

Age
& -0.011 & 0.005 & 0.019
& --- & --- & --- \\

CCI $= 2$
& -0.241 & 0.098 & 0.014
& --- & --- & --- \\

CCI $> 2$
& -0.257 & 0.121 & 0.034
& --- & --- & --- \\

BMI $\ge 30$
& -0.712 & 0.107 & $<0.001$
& 0.511 & 0.171 & 0.003 \\

Prev. hosp $\ge 1$
& -0.213 & 0.094 & 0.023
& --- & --- & --- \\

DLCO$\%$
& 0.014 & 0.002 & $<0.001$
& -0.016 & 0.004 & $<0.001$ \\

PIM$\%$
& 0.005 & 0.002 & 0.006
& --- & --- & --- \\

6MWT $\ge 460$
& 0.264 & 0.121 & 0.029
& -1.000 & 0.162 & $<0.001$ \\

FEV1$\%$ $<50$
& --- & --- & ---
& -0.320 & 0.168 & 0.056 \\

\bottomrule
\end{tabular}
\end{table}

\vspace{0.5cm}
\noindent Overall, the three models identified a remarkably consistent set of determinants of physical activity. Older age, greater comorbidity burden, obesity, and previous hospitalizations were associated with less favourable activity outcomes, whereas better pulmonary function (particularly DLCO\%), greater inspiratory muscle strength, and better exercise capacity (6MWT) were associated with higher levels of physical activity and, in some cases, lower variability and a reduced probability of not doing bouts.

\subsection{Survival probability risk estimation}\label{subsec:results_survival}
We fitted Cox proportional hazards models using the GAMLSS-derived quantities as described in Section~\ref{subsec:methods_survival}. For semi-continuous outcomes, namely $\text{Time}_{bd}$ and $\text{METarea}_{bd}$, we considered three components: the estimated probability of doing zero bouts, $\widehat{\nu}$, together with the conditional mean and standard deviation of the positive responses, denoted by $\widehat{\mu}$ and $\widehat{\mathrm{Sd}}$, respectively. For the outcome $\text{Count}_{b}$, only $\widehat{\mu}$ and $\widehat{\mathrm{Sd}}$ were included. Note that the estimated probability of zero bouts is the same when modelling $\text{Time}_{bd}$ and $\text{METarea}_{bd}$ outcomes.

The hazard function for semi-continuous outcomes is therefore specified as
\begin{equation*}\label{eq:results.surv.minutes}
   h(t \mid \widehat{\nu}, \widehat{\mu}, \widehat{\mathrm{Sd}})
=
h_0(t)\exp\left(
\omega_{\nu} \widehat{\nu}
+
\omega_{\mu} \widehat{\mu}
+
\omega_{\mathrm{Sd}} \widehat{\mathrm{Sd}}
\right),
\end{equation*}

whereas for $\text{Count}_{b}$, the hazard function is given by
\begin{equation*}\label{eq:results.surv.bouts}
   h(t \mid \widehat{\mu}, \widehat{\mathrm{Sd}})
=
h_0(t)\exp\left(
\omega_{\mu} \widehat{\mu}
+
\omega_{\mathrm{Sd}} \widehat{\mathrm{Sd}}
\right).
\end{equation*}

Table~\ref{barrio:tab_survival_models} summarizes the numerical results obtained when fitting the Cox models. For $\text{Time}_{bd}$, the mean component was a strong predictor of mortality, with a hazard ratio (HR) of 0.94 (95\% CI: 0.91--0.97), indicating that higher expected time in bouts is associated with lower risk. The standard deviation showed a positive association with mortality (HR = 1.03, 95\% CI: 1.00--1.06), suggesting that greater variability in bout duration may be detrimental, although this effect was only marginally significant. The zero component also exhibited a positive association (HR = 5.24, 95\% CI: 0.73--37.64), indicating that a higher probability of not engaging in bouts is associated with increased risk, although this effect did not reach statistical significance at the 5\% level.

For the $\text{Count}_{b}$, only the mean component was statistically significant (HR = 0.89, 95\% CI: 0.85--0.94), while the variability was not associated with mortality (HR = 1.06, 95\% CI: 0.97--1.16), indicating that the frequency of bouts is prognostic, but its dispersion is not.

For $\text{METarea}_{bd}$, all three components were statistically significant predictors of mortality. The mean was negatively associated with risk (HR = 0.99, 95\% CI: 0.98--0.996), while both the variability (HR = 1.01, 95\% CI: 1.002--1.013) and the probability of zero activity (HR = 13.41, 95\% CI: 1.63--110.14) were positively associated with mortality. These results suggest that not only the amount but also the regularity and the fact of doing or not doing bouts are relevant for survival.

Finally, we fitted a joint Cox model combining the patient-specific quantities derived from the different physical activity measures. When including the probability of zero bouts and the mean and variability of $\text{Time}_{bd}$, together with the mean number of bouts, only the components related to minutes remained associated with mortality. In particular, both the mean (HR = 0.92, 95\% CI: 0.88--0.97) and the standard deviation (HR = 1.05, 95\% CI: 1.01--1.09) of minutes in bouts were significant predictors, and the probability of zero bouts also showed a positive association with risk (HR = 13.72, 95\% CI: 1.34--140.8). In contrast, the mean number of bouts was not statistically significant after adjustment (HR = 1.05, 95\% CI: 0.98--1.12).

Overall, these findings suggest that survival is primarily driven by whether patients do bouts, the total time accumulated in such activity, and its consistency. Increasing the duration of activity bouts appears to be more relevant for improving survival, particularly when such activity is performed regularly.

\begin{table}[!ht]
\centering
\caption{Cox proportional hazards models based on GAMLSS-derived components for physical activity measures.}
\label{barrio:tab_survival_models}
\begin{tabular}{lrrrrr}
\toprule
& Est. & Std. Err & HR & p-value & c-index\\
\midrule
\multicolumn{6}{l}{\textbf{Minutes in bouts per day} ($\text{Time}_{bd}$)} \\

$\nu$ (zero probability)
& 1.656 & 1.006 & 5.241 & 0.100 & 0.772 \\

$\mu$ (mean)
& -0.058 & 0.016 & 0.943 & $<0.001$ & \\

Sd (variability)
& 0.030 & 0.016 & 1.030 & 0.062 & \\

\midrule
\multicolumn{6}{l}{\textbf{Total number of bouts} ($\text{Count}_{b}$)} \\

$\mu$ (mean)
& -0.113 & 0.026 & 0.893 & $<0.001$ & 0.724 \\

Sd (variability)
& 0.057 & 0.046 & 1.059 & 0.215 &\\

\midrule
\multicolumn{6}{l}{\textbf{Area under MET per day} ($\text{METarea}_{bd}$)} \\

$\nu$ (zero probability)
& 2.596 & 1.074 & 13.410 & 0.016 & 0.735\\

$\mu$ (mean)
& -0.010 & 0.003 & 0.990 & $<0.001$ & \\

Sd (variability)
& 0.007 & 0.003 & 1.007 & 0.014 & \\

\midrule
\multicolumn{5}{l}{\textbf{Joint model (minutes + bouts)}} \\

$\nu$ (minutes)
& 2.619 & 1.188 & 13.723 & 0.027 & 0.769\\

$\mu$ (bouts)
& 0.048 & 0.035 & 1.049 & 0.167 & \\

$\mu$ (minutes)
& -0.081 & 0.023 & 0.922 & $<0.001$ & \\

Sd (minutes)
& 0.044 & 0.019 & 1.045 & 0.022 & \\

\bottomrule
\end{tabular}
\end{table}

\subsection{Interpretation of the joint effect of the mean and variability}\label{subsec:results_interpretation}

\begin{figure}[!ht]\centering
\includegraphics[width=0.9\textwidth]{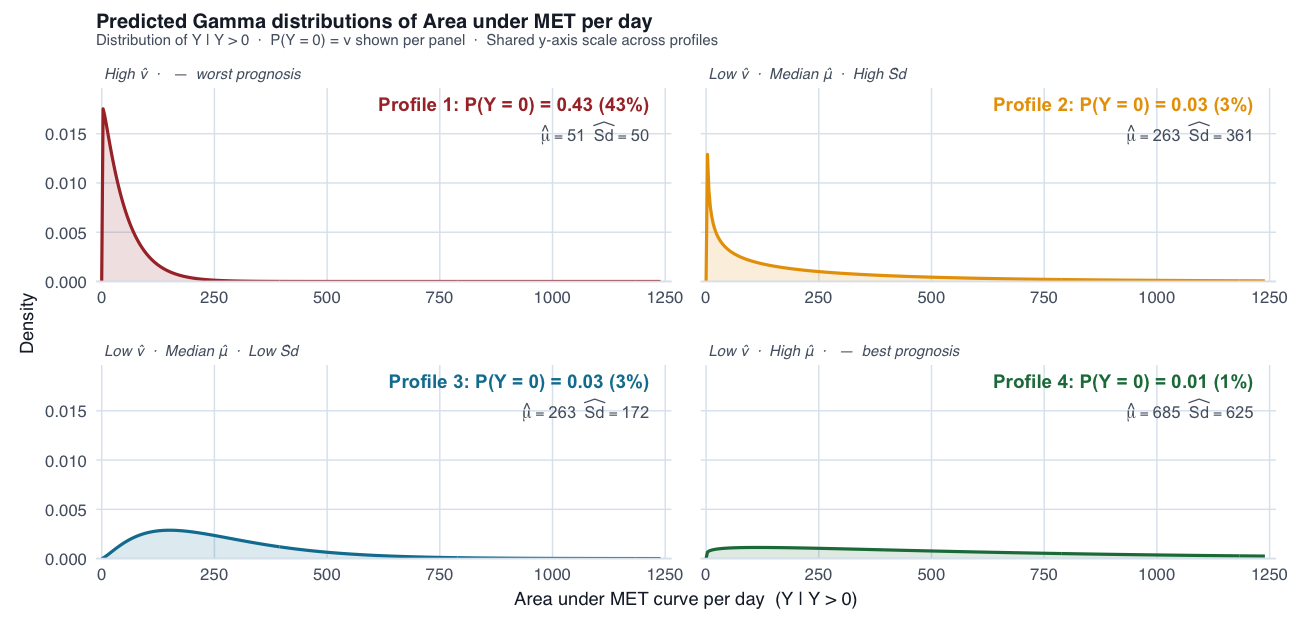}
\caption{\label{fig:profiles_area} Predicted ZAGA distributions of area under MET per day for four representative COPD patient profiles derived from the GAMLSS model.}
\end{figure}

To facilitate interpretation of the joint effects of the different parameters of the distribution of PA activity outcomes in patient's survival, we defined four representative patient profiles based on the covariates significantly associated with the zero-adjustment component ($\nu$), the mean ($\mu$) and the scale ($\sigma$) parameters of the estimated GAMLSS model. In particular, we considered the model shown in Table~\ref{tab:METarea_ZAGA}. Specifically, the profiles were defined using representative combinations of the covariates contributing to the location ($\mu$), scale ($\sigma$), and zero-adjustment ($\nu$) components of the fitted model. Continuous covariates were assigned values below, close to, or above their sample median (Table~\ref{tab:descriptive}), whereas categorical covariates were set to clinically meaningful categories. The selected combinations generated four profiles spanning a broad range of expected PA levels, variability, and probability of inactivity. 
The predicted ZAGA distributions obtained for these profiles are shown in Figure~\ref{fig:profiles_area}. Importantly, the profiles were not only characterised by differences in the mean PA, but also by differences in the variability parameter, allowing us to assess whether patients with similar expected PA levels but distinct activity patterns exhibited different survival experiences.

\begin{figure}[!ht]\centering
\includegraphics[width=0.65\textwidth]{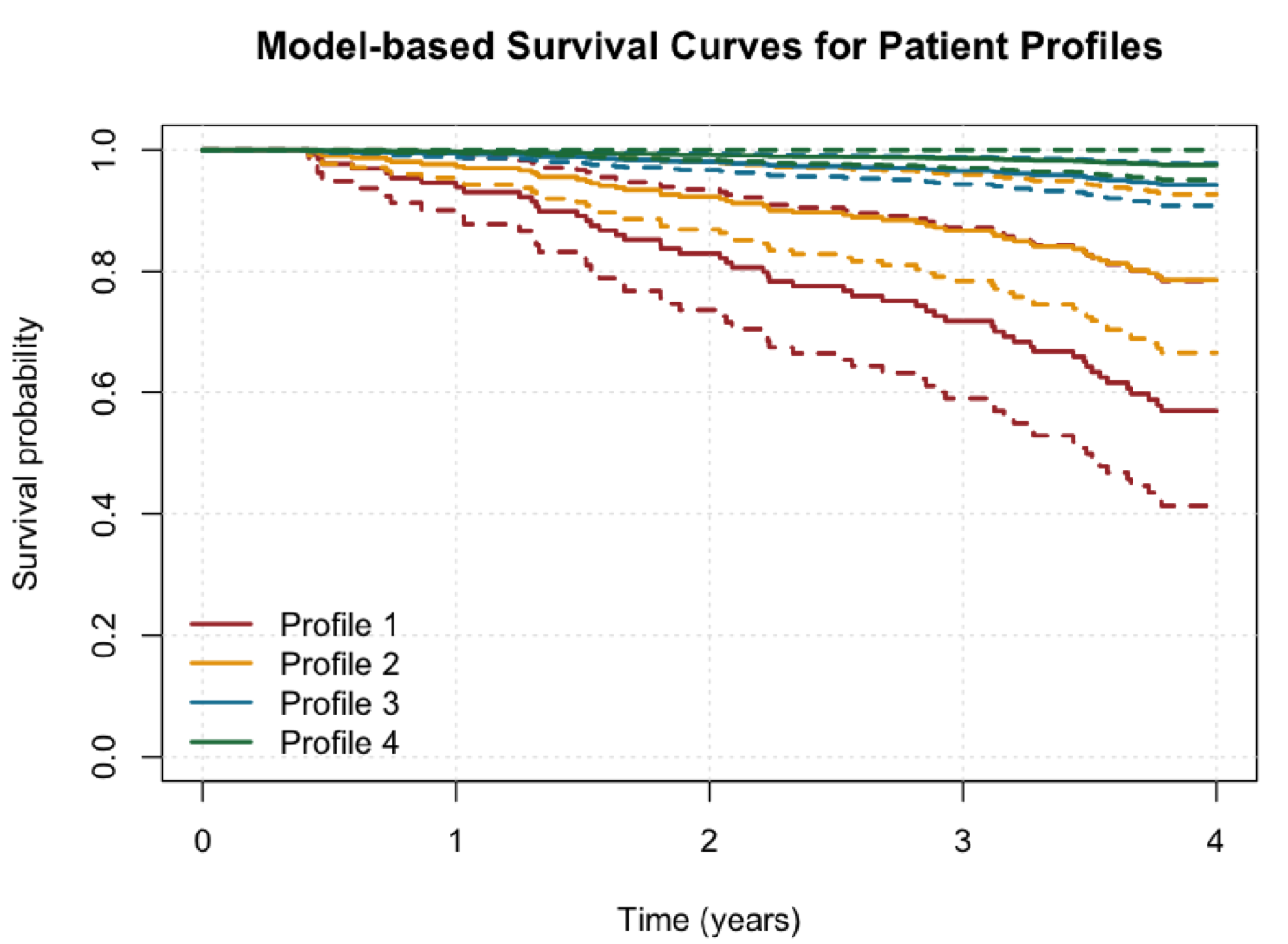}
\caption{\label{fig:profiles_survival} Model-based Survival Curves for the considered profiles.}
\end{figure}

Figure~\ref{fig:profiles_survival} further illustrates the value of modelling the full distribution of PA outcomes. In particular, Profiles 2 and 3 were constructed to have the same expected level of PA but different variability. Despite this similarity in the mean response, the two profiles exhibited distinct survival risks, indicating that the dispersion parameter captures clinically relevant information that would be overlooked by approaches focusing exclusively on the mean. These findings support the use of GAMLSS-based modelling, as they suggest that variability in physical activity may contain information relevant to survival beyond that conveyed by average activity levels alone.

\section{Discussion}

In this work, we have proposed a new framework for modelling physical activity data collected from accelerometers in patients with COPD. We have shown how an advanced modelling strategy, in which several parameters of the outcome distribution are modelled rather than focusing exclusively on the mean, can provide clinically relevant information. To the best of our knowledge, most studies investigating physical activity in patients with chronic diseases have focused on average levels of activity and have not considered different distributional characteristics of the outcomes of interest \citep{elmesmari2017accelerometer, barker2019physical, hall2020systematic, huang2024objectively}.

The relevance of GAMLSS in clinical research is not new. Several studies have applied this methodology and highlighted the importance of modelling distributional parameters beyond the mean \citep{rigby2005generalized,stasinopoulos2018gamlss}. However, in the present work, we go one step further by showing how the different parameters estimated through GAMLSS can be related to survival outcomes.

The results obtained in this study should not be interpreted as evidence of causality. Rather, they illustrate how different parameters of the distribution of physical activity outcomes may be related to patient survival and how these parameters can be described through clinically relevant variables.

In particular, patients with similar average physical activity levels may have different survival experiences when differences in activity variability are taken into account. In our analysis, profiles with comparable expected activity levels but distinct variability patterns exhibited different survival trajectories, suggesting that variability may capture aspects of physical activity behaviour that are not reflected by the mean alone. This is in line with previous research in other chronic conditions. For instance, in patients with heart failure, the skewness of accelerometer-derived activity patterns has been shown to predict all-cause mortality independently of established prognostic scores and of the mean activity level itself \citep{melin2016variability}. From a clinical perspective, this suggests that characterising physical activity exclusively through average levels may overlook relevant information contained in activity patterns.

These findings support the use of GAMLSS-based modelling in this context. By explicitly accounting for variability in physical activity outcomes, the proposed framework revealed differences in survival that would not have been apparent from the mean response alone.

Despite these promising findings, several limitations should be considered when interpreting the results. On the one hand, it is important to note that the conclusions drawn from this work are based on a cohort of patients with COPD recruited from a single hospital. Therefore, the findings cannot be assumed to be directly generalizable to the general population, to patients with other chronic conditions, or to healthcare settings in other countries. Although we believe that some of the associations observed between physical activity patterns and survival may also be present in other contexts, further studies are needed to investigate their reproducibility and generalizability.

Another limitation concerns the outcomes considered in the analysis, all of which were based on the concept of a bout. The definitions adopted in this work were derived from criteria previously used in the literature \citep{donaire2013}, defining a bout as a minimum of 10 consecutive minutes at an intensity above 3 METs. Notably, we did not allow for brief interruptions within a bout. Previous work has shown that allowing versus disallowing short interruptions materially affects the estimated frequency and duration of activity bouts \citep{ayabe2014interruption}, suggesting that alternative definitions of a bout could lead to different results. In addition, the choice of which distributional parameters of physical activity to investigate is inherently an analytical decision.

On the other hand, physical activity was assessed using accelerometer data collected during the first week of follow-up. The duration of accelerometer monitoring was short compared with the overall follow-up period of the study, resulting in a mismatch between the period during which physical activity was measured and the period over which survival was observed. Consequently, it remains unclear to what extent physical activity recorded during a single week reflects patients' long-term activity patterns over approximately four years of follow-up. Bridging the gap between the relatively short period of physical activity monitoring and the much longer clinical follow-up remains an open methodological challenge.

In addition, rather than analysing the longitudinal accelerometer MET data directly, we defined a set of summary outcomes derived from these measurements. Although such summaries facilitate interpretation and modelling, they do not fully exploit the rich temporal information available in the original accelerometer data. This limitation has also been highlighted by \citet{matabuena2023distributional}, who argued that traditional summary measures derived from accelerometer data may discard relevant information contained in the original activity measurements and proposed an alternative framework based on the distributional representation of the complete accelerometer signal. Consequently, potentially relevant temporal patterns of physical activity may be overlooked, and any measurement error associated with the derived outcomes may propagate through the different stages of the modelling process and the estimation of the distributional parameters. Nevertheless, these limitations highlight several challenges and opportunities for future research on modelling accelerometer data. In particular, they motivate the joint analysis of longitudinal accelerometer data and survival outcomes. Rather than studying the distributional parameters of PA-related variables in a first step and analyzing their relationship with survival outcomes afterwards, future work could explore approaches that make direct use of the longitudinal information collected by these devices. More generally, developing statistical methods capable of appropriately integrating information collected from devices that record physical activity intermittently over time, for example, one week of monitoring per year of follow-up, represents an interesting and clinically relevant area for further investigation.

Finally, a further extension of the present work would be the joint modelling of the different physical activity outcomes considered in this study. By definition, measures such as time accumulated in bouts, area under the MET curve during bouts, and number of bouts are strongly related, and modelling their joint distribution through copula-based approaches is a promising direction that we are currently exploring. Copula-based extensions of the GAMLSS framework, which combine flexible marginal distributional models with a copula to capture the dependence structure between outcomes, have already been proposed for bivariate and multivariate responses \citep{marra2017bivariate, espasandin2018bivariate, klein2015bayesian, kock2025truly} and could offer a natural way to jointly characterise the distribution of several physical activity outcomes within a single coherent framework.

To sum up, the results presented in this work indicate that relevant information may be contained not only in the average level of physical activity but also in other characteristics of its distribution, particularly its variability, which can jointly explain different survival trajectories of COPD patients.

\section*{Acknowledgements}
Special thanks to the Alexander von Humboldt Foundation, whose scholarship made this work possible and helped establish the network needed to continue working on the future lines of research outlined here.

\section*{Declaration of conflicts of interest}
The authors declared no potential conflicts of interest with respect to the research, authorship, and/or publication of this article.

\section*{Funding}
This work was supported in part by the Department of Education, Linguistic Policy and Culture of the Basque Government [MATHMODE Group IT1866-26, BERC 2022-2025]; the Ministry of Science and Innovation [CEX2021-001142-S / MICIN / AEI / 10.13039/501100011033, MICIU/AEI/10.13039/2501100011033, PID2024-156800OB-I00, RED2024-153680-T (BIOSTATNET)]; by the Instituto de Salud Carlos III (ISCIII) through the project RD24/0005/0020 (Red de Investigaci\'on en Cronicidad, Atenci\'on Primaria y Prevenci\'on y Promoci\'on de la Salud) and by the Alexander von Humboldt Stiftund.

\section*{Data Availability}
For data protection reasons, the data used in this study cannot be made public; however, it can be provided upon reasonable request to the authors.

\bibliographystyle{apalike}
\bibliography{references}

\end{document}